\documentclass[a4paper,twocolumn,english,aps,prl,superscriptaddress]{revtex4-1}
\usepackage[T1]{fontenc}
\usepackage[latin9]{inputenc}
\usepackage{color}
\usepackage{babel}
\usepackage{amsmath}
\usepackage{amssymb}
\usepackage{graphicx}
\usepackage{esint}
\usepackage[unicode=true,pdfusetitle,
 bookmarks=true,bookmarksnumbered=false,bookmarksopen=false,
 breaklinks=false,pdfborder={0 0 1},backref=false,colorlinks=false]
 {hyperref}

\makeatletter

\usepackage{babel}

\makeatother

\begin{document}
\renewcommand{\figurename}{FIG.} 
\title{Exceptional Andreev spectrum and supercurrent in $p$-wave non-Hermitian
Josephson junctions }
\author{Chang-An Li}
\affiliation{Institute for Theoretical Physics and Astrophysics, University of
Würzburg, 97074 Würzburg, Germany}
\affiliation{Würzburg-Dresden Cluster of Excellence ct.qmat, Germany}
\author{Björn Trauzettel}
\affiliation{Institute for Theoretical Physics and Astrophysics, University of
Würzburg, 97074 Würzburg, Germany}
\affiliation{Würzburg-Dresden Cluster of Excellence ct.qmat, Germany}
\date{\today}
\begin{abstract}
We investigate the spectrum of Andreev bound states and supercurrent
in a $p$-wave non-Hermitian Josephson junction (NHJJ) in one dimension.
The studied NHJJ is composed of two topological $p$-wave superconductors
connected by a non-Hermitian dissipative junction. Starting from the
effective non-Hermitian Bogoliubov-de Gennes bulk Hamiltonian, we
find that a pair of exceptional points emerge in the complex spectrum
of Andreev quasi-bound states. The two exceptional points with zero
energy locate symmetrically with respect to Josephson phase difference
$\phi=\pi$, at which a Majorana zero mode persists. Notably, the
exceptional points descend from a pair of Majorana zero modes after
turning on the non-Hermiticity and are topologically protected. By
analyzing the non-Hermitian scattering process at the junction, we
explicitly demonstrate the loss of quasiparticles through the decay
of scattering amplitude probabilities. Furthermore, we obtain the
supercurrent directly by the inelastic Andreev reflection amplitudes,
which provides a more intuitive interpretation of transport properties
in NHJJs. The supercurrent varies continuously as a function of $\phi$
across the exceptional points. No enhancement of critical current
is observed. We also generalize our analysis to a mixed $s$-$p$
wave NHJJ. Our results provide new insights on transport properties
of Josephson junctions in presence of Majorana zero modes, exceptional
points, and non-Hermiticity.
\end{abstract}
\maketitle

\section{Introduction}

The Josephson effect is one of the most celebrated macroscopic quantum
phenomena in condensed matter physics\ \citep{Josephson_62}. It
originates from the quantum coherence of superconductors. In presence
of a superconducting phase difference across the Josephson junction,
which can happen if two superconductors are separated by a (weak)
link, a supercurrent flows without any bias voltages\ \citep{Likharev79rmp,Beenakker92proceed,Golubov04rmp,Tinkham}.
In the short junction regime, the supercurrent is entirely determined
by the low-energy Andreev spectrum\ \citep{Beenakker91prl2,Furusaki99sm,Kwon04epj,FuL09prb,Beenakker13prl,Dolcini15prb}.
When the two superconductors are made of $p$-wave superconductors,
the resulting topological Josephson junction exhibits Majorana zero
modes and characteristic $4\pi$-periodic current-phase relations\ \citep{Alicea12rpp,Aguado17review}.

Recently, there has been growing interest in the non-Hermitian Josephson
effect\ \citep{LiCA24prb,Cayao24prb}. The non-Hermiticity arising
from coupling to the environment renders particular Andreev spectrum
and transport features of Josephson junctions\ \citep{LiCA24prb,Cayao24prb,ShenPX24prl,Beenakker24apl,Aguado25prb,Cayao25prb3,Ohnmacht25prl,Cayao24prb2,Capecelatro25prb,Sten25arxiv,Solow25arxiv,Ogino25arxiv,VanWees91prb}.
Different from the Hermitian case, the spectrum of Andreev bound states
in NHJJ becomes complex in general. The complex-valued nature of the
spectra gives rise to unique features such as exceptional points (EPs)
and exotic transport properties\ \citep{Dembowski01prl,Kawabata19prl,Bergholtz21rmp}.
However, for NHJJ in which the superconductors are of $p$-wave pairing
order, its spectral features are not fully explored and the transport
properties in presence of exceptional points are under debate.

In this work, we study the low-energy Andreev spectrum and current-phase
relation of a $p$-wave short NHJJ. The NHJJ is consisted of two $p$-wave
superconductors connected by a non-Hermitian barrier {[}Fig. \ref{fig1:setup}(a){]}.
By solving the non-Hermitian Bogoliubov-de Gennes equation, we derive
the spectral properties of Andreev quasi-bound states. We find that
a pair of EPs appear with zero real energy in the complex Andreev
spectrum {[}Fig. \ref{fig1:setup}(b){]}. Effectively, they evolve
from a pair of Majorana zero modes (MZMs) as varying the non-Hermiticity
and are topologically protected. The scattering process at the junction
turns out to be inelastic due to the non-Hermitian barrier. By analyzing
the inelastic Andreev reflection amplitudes, we can obtain the dc
supercurrent flowing across the junction directly. The resulting current-phase
relation shows continuous and smooth behavior across the EPs. No enhancement
of critical current is found.

The rest of the article is organized as follows. In Sec. II, we present
the Bogoliubov-de Gennes Hamiltonian of the $p$-wave NHJJ. In Sec.
III, we demonstrate the properties of EPs in Andreev spectra and their
topological protection. In Sec. IV, we calculate the Andreev and normal
scattering amplitudes across the NHJJ. In Sec. V, we obtain the supercurrent
from the inelastic Andreev reflection amplitudes. In Sec. VI, we generalize
our discussion to the mixed $s$-$p$ wave NHJJ. Finally, we conclude
in Sec. VII.

\section{Model and particle-hole symmetry}

\begin{figure}
\includegraphics[width=1\linewidth]{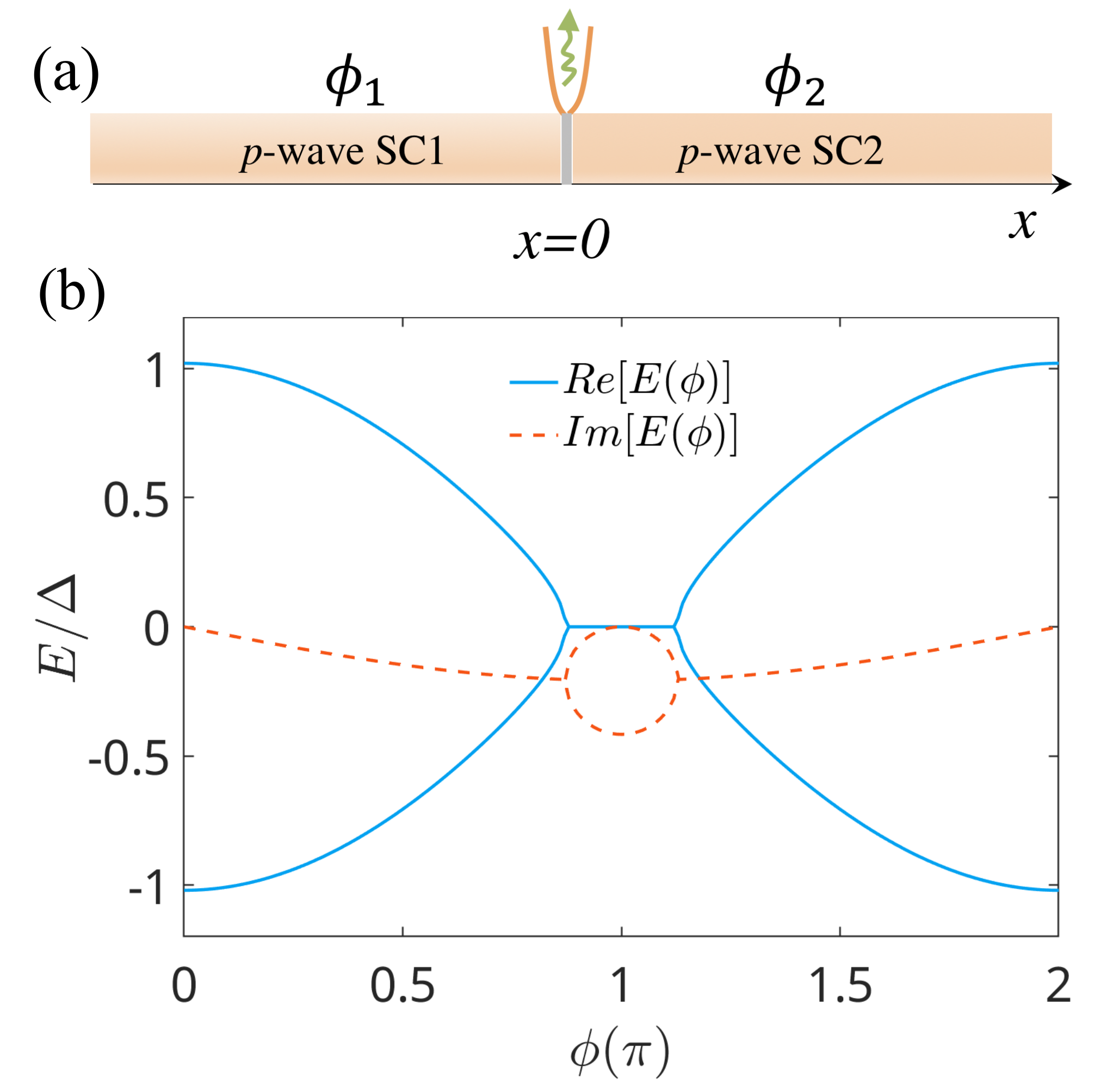}

\caption{(a) Sketch of a 1D $p$-wave NHJJ: two triplet $p$-wave superconductors
are connected by a short junction, which is coupled to the environment.
The two superconductors possess a phase difference $\phi=\phi_{2}-\phi_{1}$.
(b) Andreev spectrum as a function of $\phi$ corresponding to the
setup in (a). A pair of exceptional points appear, implying unique
non-Hermitian spectral features. We choose $Z=0.2$ for the non-Hermitian
barrier strength. \label{fig1:setup}}
\end{figure}

We study the properties of a one-dimensional (1D) $p$-wave NHJJ sketched
in Fig. \ref{fig1:setup}(a). The junction is coupled to an external
fermionic environment at the junction, constituting an open quantum
system. The Bogoliubov-de Gennes (BdG) Hamiltonian describing the
NHJJ can be written as

\begin{equation}
H_{\mathrm{BdG}}=\left(\begin{array}{cc}
-\frac{\hbar^{2}\partial_{x}^{2}}{2m}-\mu+U(x) & \hat{\Delta}(x)\\
\hat{\Delta}^{\dagger}(x) & \frac{\hbar^{2}\partial_{x}^{2}}{2m}+\mu-U^{*}(x)
\end{array}\right),\label{eq:BdGHamiltonian}
\end{equation}
where $m$ is the effective mass of the electrons, $\mu$ the chemical
potential, and $\hat{\Delta}(x)$ the pairing potential chosen as
triplet $p$-wave type. We have previously justified this effective
non-Hermitian Hamiltonian description based on the Lindbladian formalism
under certain approximations\ \citep{LiCA24prb}. The coupling to
environment gives rise to an imaginary potential at the junction as
$U(x)=-iV\delta(x)$ where $V>0$. This term represents the continuous
loss of quasiparticles from the system into the environment\ \citep{Breuer02book,Nazarovbook,Daley14AIP,Minganti19pras}.

The particle-hole symmetry of the type $UH_{\mathrm{BdG}}^{*}U^{-1}=-H_{\mathrm{BdG}}$
is physically more relevant in NHJJs\ \citep{LiCA24prb}, where $U$
is a unitary matrix. In literatures, this type of particle-hole symmetry
is labeled as $\mathrm{PHS}^{\dagger}$\ \citep{Kawabata19prl,Kawabata19prx}.\textcolor{blue}{{}
}The electron and hole excitations are governed by the BdG equation
$H_{\mathrm{BdG}}\psi(x)=E\psi(x)$ where $\psi(x)$ is the eigenstate
and $E$ denotes the excitation energy measured relative to the Fermi
energy. Due to $\mathrm{PHS}^{\dagger}$, the energy spectrum of $H_{\mathrm{BdG}}$
exhibits a symmetry: for every eigenenergy $E$, there exists a corresponding
eigenenergy $-E^{*}.$

\section{Exceptional Andreev spectrum}

Without loss of generality, we assume that the left and right superconductors
of the $p$-wave NHJJ have a pairing potential of the same magnitude
but different phases as\ \citep{Kwon04epj}

\begin{align}
\hat{\Delta}(x) & =\begin{cases}
\Delta e^{i\phi_{1}}\hat{k}_{x}/k_{F}, & x<0;\\
\Delta e^{i\phi_{2}}\hat{k}_{x}/k_{F}, & x>0,
\end{cases}
\end{align}
where $\phi=\phi_{2}-\phi_{1}$ is the phase difference across the
junction and $k_{F}$ the Fermi wavevector. The phase difference leads
to the low-energy bound states existing at the junction.\textcolor{blue}{{}
}We obtain such Andreev quasi-bound states in the NHJJ by solving
the BdG equation

\begin{equation}
H_{\mathrm{BdG}}\left(\begin{array}{c}
u(x)\\
v(x)
\end{array}\right)=E\left(\begin{array}{c}
u(x)\\
v(x)
\end{array}\right),\label{eq:BdGEquation}
\end{equation}
where the eigenstate $\psi(x)=(u(x),v(x))^{T}$ is a mixture of electron
and hole components\ \citep{Furusaki91ssc,Beenakker91prl}. 

We choose trial wavefunctions for different regions. For $x<0$, it
takes 
\begin{equation}
\psi(x<0)=ae^{ik_{h}x}\left(\begin{array}{c}
v_{0}\\
u_{0}
\end{array}\right)+be^{-ik_{e}x}\left(\begin{array}{c}
u_{0}\\
-v_{0}
\end{array}\right),
\end{equation}
and for $x>0$, it takes 
\begin{equation}
\psi(x>0)=ce^{ik_{e}x}\left(\begin{array}{c}
u_{0}e^{i\phi/2}\\
v_{0}e^{-i\phi/2}
\end{array}\right)+de^{-ik_{h}x}\left(\begin{array}{c}
v_{0}e^{i\phi/2}\\
-u_{0}e^{-i\phi/2}
\end{array}\right),
\end{equation}
where $u_{0}^{2}=\frac{1}{2}\left(1+i\frac{\sqrt{\Delta^{2}-E^{2}}}{E}\right),v_{0}^{2}=\frac{1}{2}\left(1-i\frac{\sqrt{\Delta^{2}-E^{2}}}{E}\right)$
and $\hbar k_{e/h}=\sqrt{2m(\mu\pm i\sqrt{\Delta^{2}-E^{2}})}$. In
the following calculations, we employ the Andreev approximation where
the velocity of quasiparticles are set to be equal to the Fermi velocity.
The coefficients $a,b,c,d$ are determined by the continuity of wavefunctions
and the corresponding derivatives at the boundary as 
\begin{align}
\psi(0+) & =\psi(0-),\nonumber \\
\partial_{x}\psi(0+)-\partial_{x}\psi(0-) & =-\tau_{z}\frac{2miV}{\hbar^{2}}\psi(0-).\label{eq:Boundarycondition-1}
\end{align}
Here, $\tau_{z}$ is the Pauli matrix. In the following, we define
the parameter $Z\equiv\frac{mV}{\hbar^{2}k_{F}}$, which characterizes
the non-Hermitian potential strength, and we focus on the weak coupling
limit with $Z\ll1$.

\subsection{Majorana zero modes at $\phi=\pi$}

There exist a pair of MZMs in the Hermitian topological $p$-wave
Josephson junction at phase difference $\phi=\pi$. We find that MZMs
survive even in the non-Hermitian environment. To verify this, we
first consider the $E=0$ case in the BdG equation Eq.~\eqref{eq:BdGEquation}.
Under this condition, the basis function takes the form $\left(\begin{array}{c}
u(x)\\
v(x)
\end{array}\right)\propto e^{ixk_{e/h}}\left(\begin{array}{c}
u_{0}\\
v_{0}
\end{array}\right)=e^{ixk_{e/h}}\left(\begin{array}{c}
i\\
1
\end{array}\right)$. Notably, this wavefunction satisfies the Majorana condition $(u,v)^{T}\propto(v^{*},u^{*})^{T}$.
Next, imposing the continuity condition at the junction leads to the
secular equation 
\begin{equation}
u_{0}^{4}(Z-1)^{2}+v_{0}^{4}(Z+1)^{2}+2u_{0}^{2}v_{0}^{2}(Z^{2}-\cos\phi)=0.
\end{equation}
For $\phi=\pi$, with the choice $u_{0}=i$ and $v_{0}=1$, this equation
holds, demonstrating that a MZM persists in the $p$-wave NHJJ.

\subsection{Exceptional Andreev quasi-bound states}

For the general solution of the Andreev spectrum, a secular equation
determined from Eq.~\eqref{eq:Boundarycondition-1} is obtained as

\begin{equation}
(Z^{2}+1)+2iZ\sqrt{\frac{\Delta^{2}}{E^{2}}-1}=\Delta^{2}\frac{\cos^{2}(\frac{\phi}{2})}{E^{2}}.\label{eq:pwavesecular}
\end{equation}
At $Z=0$, the corresponding Andreev spectrum reduces to the conventional
result $E^{\pm}(\phi)=\pm\Delta\cos(\frac{\phi}{2})$\ \citep{Kwon04epj}.
The general solution of the secular equation with nonzero $Z$ yields

\begin{equation}
\frac{E^{\pm}(\phi)}{\Delta}=\frac{\pm\sqrt{\cos^{2}(\frac{\phi}{2})-Z^{2}}-iZ\sin(\frac{\phi}{2})}{1-Z^{2}}.\label{eq:Andreev spectrum}
\end{equation}
As a consequence, the Andreev quasi-bound states spectrum becomes
complex, different from the Hermitian case. The real parts characterize
the physical energy while the imaginary parts indicate a broadening
due to coupling to the environment. Interestingly, the effective broadening
of Andreev quasi-bound states is phase-dependent.

We plot the corresponding Andreev spectrum of $p$-wave NHJJs in Fig.
\ref{fig1:setup}(b). Remarkably, the complex Andreev spectrum hosts
a pair of EPs at zero real energy, which are unique to non-Hermitian
systems. The EPs are located at 
\begin{alignat}{1}
\text{\ensuremath{\phi_{\mathrm{EP}}^{1}}} & =2n\pi+2\arccos(Z),\\
\phi_{\mathrm{EP}}^{2} & =2(n+1)\pi-2\arccos(Z),
\end{alignat}
where $n\in\mathbb{Z}.$ The Andreev spectrum is complex-valued away
from the EPs while it becomes purely imaginary between the two EPs.
At the EPs, the energy takes $E_{\mathrm{EP}}=\frac{-iZ}{\sqrt{1-Z^{2}}}$
with zero real values. The separation of two EPs is $\delta\phi=2\pi-4\arccos(Z)$,
controlled by the parameter $Z$. Remarkably, we can view the appearance
of EPs as an evolution of a pair of MZMs. At $Z=0,$ the system is
Hermitian and a pair of MZMs appear at $\phi=\pi$\ \citep{Kwon04epj}.
As turning on the non-Hermiticity with finite $Z$, one of the MZMs
remains sharp (with $\mathrm{Im}(E)=0$) and the other gets broadened
(with $\mathrm{Im}(E)\neq0$). As the phase $\phi$ deviates from
$\pi$, these two modes evolve gradually while keeping zero real energy
and coalesce to form EPs at $\phi_{\mathrm{EP}}$ {[}see Fig. \ref{fig1:setup}(b){]}.
In this sense, the EPs descend from a pair of MZMs in presence of
finite $Z$. As the MZMs are topologically protected, so are the EPs.
This protection of the EPs can be understood in terms of the topological
classification of NHJJs\ \citep{Ohnmacht25prl}. The model Hamiltonian
respects particle-hole symsmetry $\mathrm{PHS}^{\dagger}$ but breaks
time reversal symmetry due to non-Hermitian dissipation \footnote{Alternatively, fixing of a Josephson phase difference generally breaks
time reversal symmetry of the junction, except at special case $\phi=\pi$}. Thus the system falls into class $\mathrm{D}^{\dagger}$\ \citep{Kawabata19prx,Ohnmacht25prl,Solow25arxiv}.
Accordingly, this class has a $\mathbb{\mathbb{Z}}_{2}$ topological
invariant in the zero-dimensional junction. We can define a topological
invariant from the vorticity of the EPs as $\nu=\frac{1}{\pi i}\oint\nabla_{\phi}\ln[E^{+}(\phi)-E^{-}(\phi)]d\phi$\ \citep{Kawabata19prl}.
The two EPs thus have opposite charges with $\nu=+1$ and $\nu=-1$.
They cannot be gapped away unless being shifted towards $\phi=\pi$
to annihilate due to particle-hole symmetry protection. These EPs
with zero real energies are stable against perturbations.

\section{Inelastic reflection and tunneling amplitudes}

Next, we analyze the scattering amplitudes of quasiparticles scattered
by the non-Hermitian barrier in the $p$-wave NHJJ. Consider an electron-like
quasiparticle that is propagating from the left side of the junction
($x<0$). After scattering at the barrier, there are two reflection
amplitudes (Andreev reflection $A_{h-}$ and normal reflection $A_{e-}$)
and two tunneling amplitudes (normal tunneling $A_{e+}$ and crossed
Andreev tunneling $A_{h+}$).\textcolor{blue}{{} }The wave function
propagating in different regions are expressed in terms of different
scattering states. For $x<0$, it can be written as

\begin{align}
\psi(x<0)= & e^{-ik_{e}x}\left(\begin{array}{c}
u_{0}\\
-v_{0}
\end{array}\right)+A_{h-}e^{ik_{h}x}\left(\begin{array}{c}
v_{0}\\
u_{0}
\end{array}\right)\nonumber \\
 & +A_{e-}e^{-ik_{e}x}\left(\begin{array}{c}
u_{0}\\
-v_{0}
\end{array}\right),
\end{align}
while for $x>0$ it reads

\begin{align}
\psi(x>0) & =A_{e+}e^{ik_{e}x}\left(\begin{array}{c}
u_{0}e^{i\phi/2}\\
v_{0}e^{-i\phi/2}
\end{array}\right)\nonumber \\
 & +A_{h+}e^{-ik_{h}x}\left(\begin{array}{c}
v_{0}e^{i\phi/2}\\
-u_{0}e^{-i\phi/2}
\end{array}\right).
\end{align}
The scattering amplitudes $A_{e\pm}$ and $A_{h\pm}$ are determined
by the boundary conditions $\psi(0+)=\psi(0-)$ and $\psi'(0+)-\psi'(0-)=-i\tau_{z}\frac{2mV}{\hbar^{2}}\psi(0+)$.
By solving the secular equation, the corresponding four scattering
amplitudes are given by 
\begin{align}
A_{h-} & =\frac{-\Delta[E(Z^{2}+\sin^{2}\frac{\phi}{2})+\Omega(Z+i\sin\frac{\phi}{2}\cos\frac{\phi}{2})]}{(Z^{2}+1)E^{2}+2Z\Omega E-\Delta^{2}\cos^{2}\frac{\phi}{2}},\label{eq:AndreevCoefficient}\\
A_{e-} & =\frac{-Z\text{\ensuremath{\Omega}}(\Omega+ZE)}{(Z^{2}+1)E^{2}+2Z\Omega E-\Delta^{2}\cos^{2}\frac{\phi}{2}},\\
A_{e+} & =\frac{\Omega E(Z\cos\frac{\phi}{2}-i\sin\frac{\phi}{2})+\Omega^{2}(\cos\frac{\phi}{2}-iZ\sin\frac{\phi}{2})}{(Z^{2}+1)E^{2}+2Z\Omega E-\Delta^{2}\cos^{2}\frac{\phi}{2}},\\
A_{h+} & =\frac{-i\Delta\Omega Z\sin\frac{\phi}{2}}{(Z^{2}+1)E^{2}+2Z\Omega E-\Delta^{2}\cos^{2}\frac{\phi}{2}},
\end{align}
where the parameter $\Omega$ is defined as $\Omega\equiv\sqrt{E^{2}-\Delta^{2}}$.
Note that the poles of the scattering amplitudes yield the Andreev
spectrum. This is clear since the denominator of these scattering
amplitudes corresponds to the secular equation Eq.~\eqref{eq:pwavesecular}
that determines the Andreev quasi-bound states.

\begin{figure}
\includegraphics[width=1\linewidth]{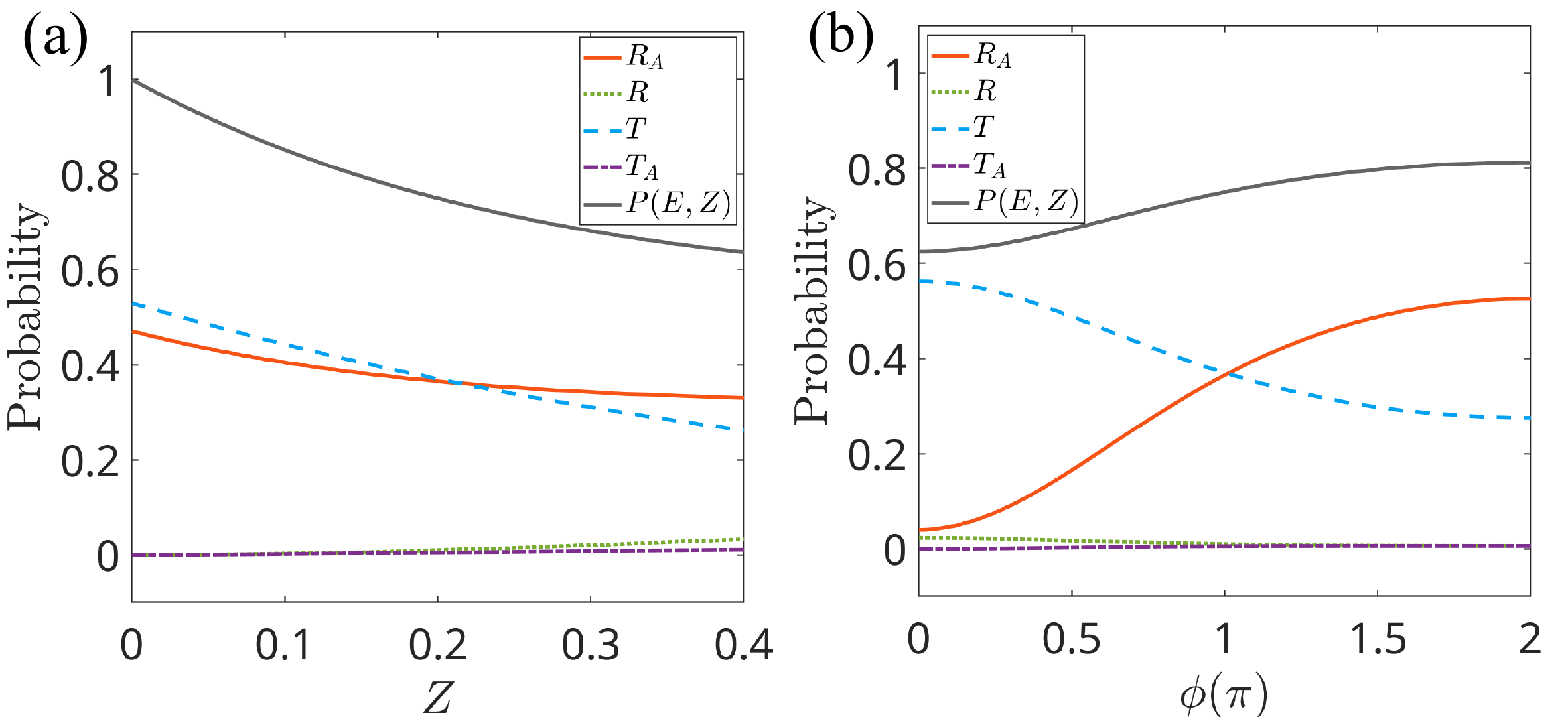}

\caption{(a) Scattering probabilities $|A_{\eta s}|^{2}$ as a function of
$Z$ with $\eta=e/h$ and $s=\pm$. $R_{A}(R)$ stands for Andreev
(normal) reflection, and $T_{A}(T)$ stands for Andreev (normal) tunneling.
We choose $\phi=\pi/2$. (b) Scattering probabilities as a function
of $\phi$ for fixed $Z=0.2$. Other parameters are chosen as $\Delta=0.2$
and $E=0.25$. \label{fig2:scatteringAmplitude}}
\end{figure}

Several important conclusions are obtained from these scattering amplitudes.
First, we find that the non-Hermitian scattering barrier leads to
a loss of quasiparticles from the junction to the environment. We
plot the scattering probabilities, i.e. $|A_{\eta s}|^{2}$ with $\eta=e/h$
and $s=\pm$, in Fig. \ref{fig2:scatteringAmplitude}(a) as a function
of $Z$. As increasing the strength of $Z$, the scattering probabilities
decrease gradually. The total scattering probability is defined as

\begin{equation}
P(E,Z)=\sum_{\eta=e/h,s=\pm}|A_{\eta s}|^{2}.
\end{equation}
In the Hermitian case at $Z=0$, the total probability is conserved
at $P=1$ as expected. However, when $Z$ is nonzero, we find that
the total probability decays continuously from $P=1$, which indicates
the loss of quasiparticles from the junction to the environment. The
scattering process in the NHJJ thus becomes inelastic. In this sense,
the value of $Z$ measures the magnitude of loss of quasiparticles.
We also note that the Andreev reflection and normal tunneling have
relatively large probabilities as compared to other scattering processes.
Moreover, we plot different scattering probabilities as a function
of $\phi$ in Fig. \ref{fig2:scatteringAmplitude}(b). Notably, these
plots change smoothly over the whole phase domain and no sudden changes
happen across the EPs.

\section{Supercurrents from inelastic Andreev reflections}

The Andreev reflection amplitudes are closely related to supercurrents
flowing across the Josephson junctions. Due to phase coherence across
the junction, quasiparticles move back and forth in the junction carrying
the phase factor of superconductors imprinted by Andreev reflection\ \citep{Tinkham,Golubov04rmp,Beenakker91prl2,Furusaki99sm}.\textcolor{magenta}{{}
}Henceforth, Andreev reflections lead to phase-coherent transport
of Cooper pairs across the junction, i.e. the dc supercurrent. The
supercurrent from Andreev reflections is given by\ \citep{Furusaki91ssc}

\begin{equation}
I(\phi)=\frac{e\Delta}{\hbar\beta}\sum_{\omega_{n}}\frac{1}{\Omega_{n}}\left[A_{h-}(\phi,i\omega_{n})-A_{h-}(-\phi,i\omega_{n})\right],\label{eq:Supercurrent1}
\end{equation}
where $\omega_{n}$ is the Matsubara frequency $\omega_{n}=(2n+1)\pi k_{B}T$
and $\beta=1/k_{B}T$ with $k_{B}$ being the Boltzmann constant and
$T$ the temperature. Here, $e$ is the electron charge and $\hbar$
is the Planck constant, and $\Omega_{n}=\sqrt{\omega_{n}^{2}+\Delta^{2}}$.
In Eq.~\eqref{eq:Supercurrent1}, $A_{h-}(\phi,i\omega_{n})$ is
the scattering amplitude which indicates a left-coming $(x<0)$ electron-like
quasiparticle is reflected as a hole-like quasiparticle. While $A_{h-}(-\phi,i\omega_{n})$
is equal to the scattering amplitude that a left-coming $(x<0)$ hole-like
quasiparticle is reflected as an electron-like quasiparticle\ \citep{Furusaki91ssc}.
Therefore, the supercurrent should be proportional to the difference
of $A_{h-}(\phi,i\omega_{n})-A_{h-}(-\phi,i\omega_{n})$. This formula
provides a more direct and intuitive interpretation of the supercurrent
from scattering point of view. Considering the same mechanism responsible
for the supercurrent, we may generalize this formula to NHJJs. The
Andreev reflection amplitude in $p$-wave NHJJ is given by Eq.~\eqref{eq:AndreevCoefficient}.
Substituting it with $E=i\omega_{n}$ into Eq.~\eqref{eq:Supercurrent1},
we then find that 
\begin{equation}
I(\phi)=-\frac{e\Delta}{\hbar\beta}\sum_{\omega_{n}}\frac{\Delta\sin(\phi)}{\omega_{n}^{2}(1+Z^{2})+2Z\Omega_{n}\omega_{n}+\Delta^{2}\cos^{2}(\phi/2)}.\label{eq:SupercurrentDenominator}
\end{equation}
We evaluate this Matsubara frequency summation by reducing it to a
contour integral. After some algebra, we arrive at 
\begin{equation}
I(\phi)=-\frac{ie\Delta^{2}}{\hbar}\frac{\sin\phi}{1+Z^{2}}\frac{f_{\mathrm{eff}}(\omega_{1})-f_{\mathrm{eff}}(\omega_{2})}{(\omega_{1}-\omega_{2})},\label{eq:Currentformula}
\end{equation}
where $f_{\mathrm{eff}}(\omega,\beta)$ is the effective Fermi-Dirac
distribution function. In non-Hermitian systems with complex eigenenergy,
it is generalized to be $f_{\mathrm{eff}}(\omega,\beta)=-\frac{1}{\pi}\left[\Psi(\frac{1}{2}+\frac{i\beta\omega}{2\pi})-\frac{i\pi}{2}\right]$
where $\Psi(z)$ is the digamma function\ \citep{ShenPX24prl}. Here,
$\omega_{1}$ and $\omega_{2}$ are two complex poles given by the
denominator from Eq.~\eqref{eq:SupercurrentDenominator} with complex
variable $z$. This result confirms that the Andreev quasi-bound states
contribute the supercurrent in the short junction limit.

Since the Andreev spectrum shows substantial difference away from
and between the two EPs, we consider the consequent supercurrent separately
for different phase domains. Away from the EPs, the Andreev spectrum
is complex-valued. Substituting the complex spectrum in Eq.~\eqref{eq:Andreev spectrum}
into the effective Fermi-Dirac distribution function, and exploiting
Eq.~\eqref{eq:Currentformula}, the corresponding supercurrent reads
\begin{equation}
I_{A}(\phi)=\frac{e\Delta(1-Z^{2})\sin\phi}{\pi\hbar(1+Z^{2})\sqrt{\cos^{2}(\frac{\phi}{2})-Z^{2}}}\mathrm{Im}\left[\Psi\left(\frac{1}{2}+\frac{i\beta E_{A}^{+}}{2\pi}\right)\right],\label{eq:currentBefore}
\end{equation}
where the phase domain is within $\phi\in[2n\pi,2n\pi+2\arccos(Z)]\cup[2(n+1)\pi-2\arccos(Z),2(n+1)\pi]$.
At $Z=0$, there are no EPs and this equation reduces to the well-known
result $I(\phi)=\frac{e\Delta}{\hbar}\sin\frac{\phi}{2}\tanh\left(\frac{\beta\Delta}{2}\cos\frac{\phi}{2}\right)$
in Hermitian Josephson junctions\ \citep{Kulik75jetps}. In contrast,
the Andreev spectrum $E_{B}^{\pm}$ becomes purely imaginary between
the EPs. In this case, the corresponding supercurrent can be directly
calculated as

\begin{align}
I_{B}(\phi) & =\frac{-e\Delta(1-Z^{2})\sin\phi}{2\pi\hbar(1+Z^{2})\sqrt{Z^{2}-\cos^{2}(\frac{\phi}{2})}}\times\nonumber \\
 & \left[\Psi\left(\frac{1}{2}+\frac{i\beta E_{B}^{+}}{2\pi}\right)-\Psi\left(\frac{1}{2}+\frac{i\beta E_{B}^{-}}{2\pi}\right)\right],\label{eq:currentAfter}
\end{align}
where the phase domain is $\phi\in[2n\pi+2\arccos(Z),2(n+1)\pi-2\arccos(Z)]$.

\begin{figure}
\includegraphics[width=1\linewidth]{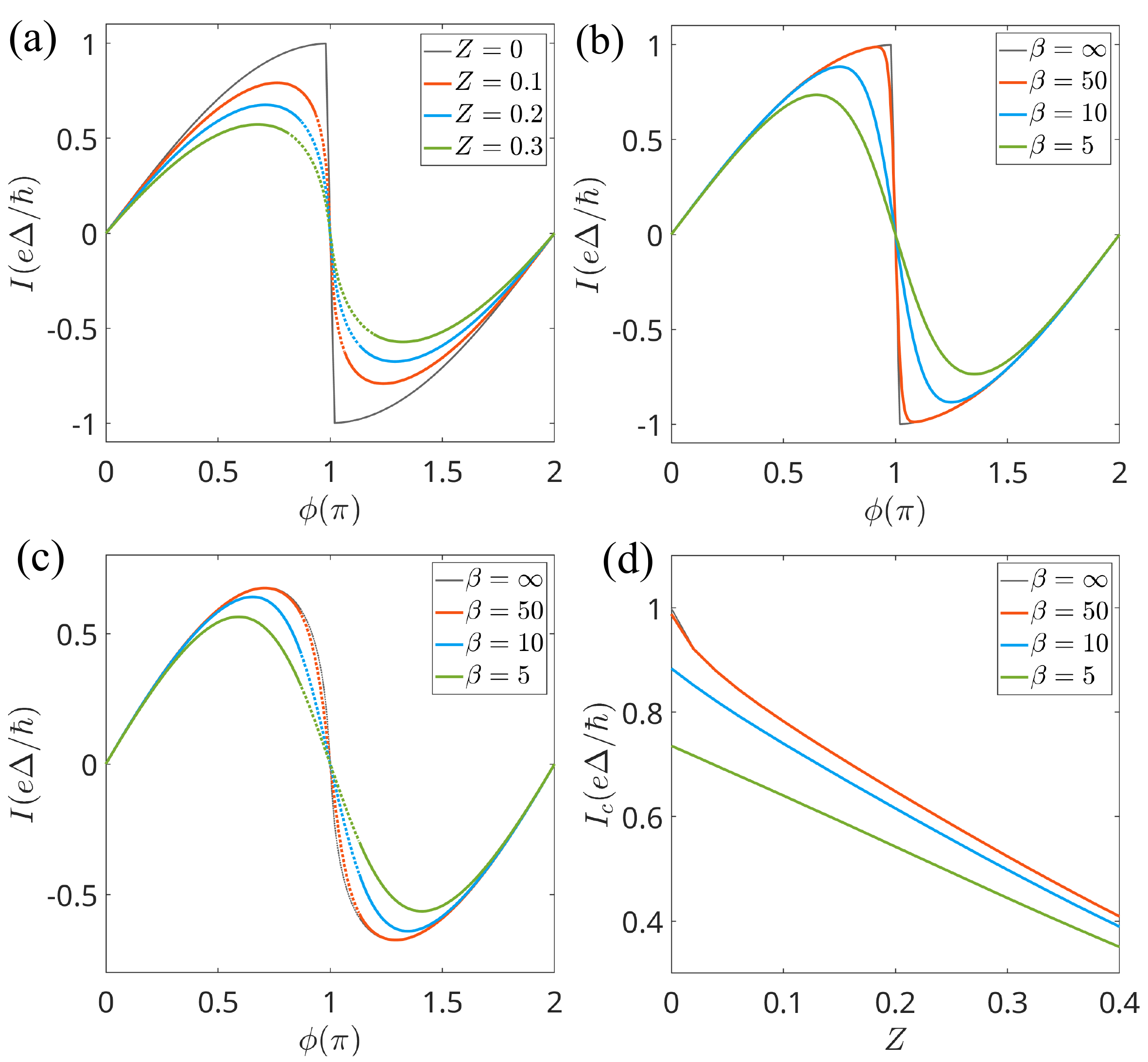}

\caption{(a) Current-phase relation for different $Z$ with temperature $T\rightarrow0$
($\beta=1/k_{B}T\rightarrow\infty$). The solid lines are from Eq.~\eqref{eq:currentBefore}
away from EPs, and dashed lines are from Eq.~\eqref{eq:currentAfter}
between EPs. (b) Current-phase relation of the Hermitian limit with
$Z=0$ at different temperatures. (c) Current-phase relation with
$Z=0.2$ at different temperatures. (d) The critical supercurrent
$I_{c}$ as a function of parameter $Z$ at different temperatures.\label{fig3:Supercurrent}}
\end{figure}

We demonstrate the current-phase relation in Figs. \ref{fig3:Supercurrent}(a,b,c)
according to Eq.~\eqref{eq:currentBefore} (solid lines) and Eq.~\eqref{eq:currentAfter}
(dashed lines) for different phase domains. We find that the supercurrent
changes smoothly across the EPs and no abrupt changes happen at the
EPs. This is consistent with previous results\ \citep{ShenPX24prl,Aguado25prb}.
From the scattering point of view, the smooth behavior of the supercurrent
originates from the continuity of wavefunctions at the junction. We
further compare the effect of temperature on current-phase relation
for $Z=0.2$ of the non-Hermitian case and $Z=0$ of the Hermitian
case, as shown in Fig. \ref{fig3:Supercurrent}(b) and Fig. \ref{fig3:Supercurrent}(c),
respectively. At low temperatures, the current-phase approaches to
a sawtooth shape in Hermitian Josephson junctions whereas it takes
a sinusoidal form in NHJJ even at zero temperature. Note that the
the critical supercurrent $I_{c}$ decreases with increasing the strength
of $Z$ {[}Fig. \ref{fig3:Supercurrent}(a) and Fig. \ref{fig3:Supercurrent}(d){]}
or with increasing temperature $T$ {[}Fig. \ref{fig3:Supercurrent}(c){]}.
We plot $I_{c}$ as a function of $Z$ in Fig. \ref{fig3:Supercurrent}(d).
We find that $I_{c}$ decreases almost linearly as increasing $Z$,
which attributes to the loss of quasiparticles from the inelastic
Andreev reflections. There is no enhancement of critical current arising
from the presence of EPs. The linear depending decay of $I_{c}$ on
$Z$ is different from the exponential decay depending on the temperature.

\section{Andreev spectrum in mixed $s$-$p$ wave non-Hermitian Josephson
junctions}

\begin{figure}
\includegraphics[width=1\linewidth]{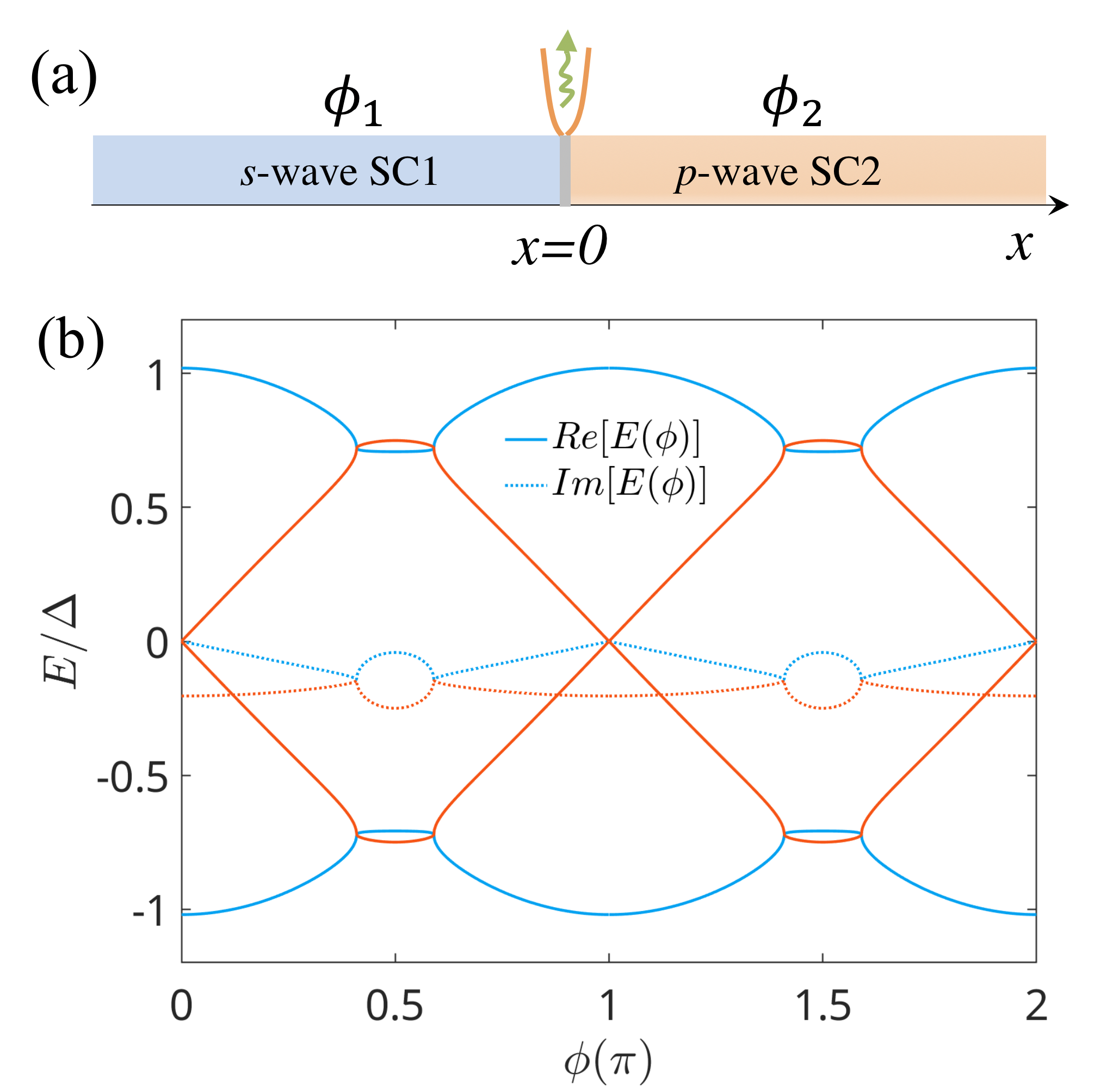}

\caption{(a) Sketch of a 1D mixed $s$-$p$ wave NHJJ: the left superconductor
is of $s$-wave paring and the right superconductor is of $p$-wave
paring. The two superconductors possess a phase difference $\phi=\phi_{2}-\phi_{1}$.
(b) Andreev spectrum as a function of $\phi$ corresponding to setup
in (a). Four pairs of exceptional points appear, implying unique non-Hermitian
spectral features. Solid lines are for the real energy and dashed
lines are for the corresponding imaginary parts. Red and blue colors
are for different pairs of Andreev spectra. We choose $Z=0.2$ in
this plot. \label{fig1:setup-2}}
\end{figure}

We further consider the properties of a mixed $s$-$p$ wave NHJJ
as shown in Fig. \ref{fig1:setup-2}(a). The junction geometry is
constructed the same as $p$-wave NHJJ in Fig. \ref{fig1:setup}(a)
but with one of superconductors replaced by an $s$-wave superconductor.
The BdG Hamiltonian describing the system takes the same form as in
Eq.~\eqref{eq:BdGHamiltonian} but with the pairing potential being
modified to

\begin{align}
\hat{\Delta}(x) & =\begin{cases}
\Delta e^{i\phi_{1}}, & x<0;\\
\Delta e^{i\phi_{2}}\hat{k}_{x}/k_{F}, & x>0.
\end{cases}
\end{align}
Following similar steps as above, we solve the BdG equation for the
mixed $s$-$p$ wave NHJJ. The corresponding secular equation is given
by
\begin{equation}
(Z^{2}+1)\sqrt{\frac{\Delta^{2}}{E^{2}}-1}-iZ(2-\frac{\Delta^{2}}{E^{2}})=\frac{\Delta^{2}\sin\phi}{2E^{2}}.\label{eq:SecularEq_spwave}
\end{equation}
Let us first check the simple case with $Z=0$, where the secular
equation reduces to $\sqrt{\frac{\Delta^{2}}{E^{2}}-1}=\frac{\Delta^{2}\sin\phi}{2E^{2}}$.
It gives rise to $E=\pm\Delta\cos\frac{\phi}{2}$ and $E=\pm\Delta\sin\frac{\phi}{2}$,
which agree with the results obtained in Ref.\ \citep{Kwon04epj}.
Solving the the secular equation for nonzero $Z$, it gives the Andreev
spectrum by
\begin{equation}
\frac{E^{2}}{\Delta^{2}}=\frac{(1-Z^{2})^{2}-2iZ\sin\phi\pm(1+Z^{2})\sqrt{(1-Z^{2})^{2}-\sin^{2}\phi}}{2(1-Z^{2})^{2}}.\label{eq:Andreev spectrum_sp}
\end{equation}
Note that there are four bands in the mixed $s$-$p$ wave NHJJ. These
Andreev spectra of quasi-bound states are illustrated in Fig. \ref{fig1:setup-2}(b).
Compared with the $p$-wave NHJJ, more EPs emerge and they are shifted
away from zero real energy. The appearance of EPs is determined by
the condition $(1-Z^{2})^{2}-\sin^{2}\phi=0$, which leads to four
pairs of EPs located at 
\begin{alignat}{1}
\text{\ensuremath{\phi_{\mathrm{EP}}^{1,4}}} & =2n\pi\pm\arcsin(1-Z^{2}),\\
\phi_{\mathrm{EP}}^{2,3} & =(2n+1)\pi\pm\arcsin(1-Z^{2}).
\end{alignat}
The corresponding energy of each EP can be determined according to
Eq.~\eqref{eq:Andreev spectrum_sp}. Importantly, the junction is
asymmetric with different parities of superconducting pairing potentials
at the two sides. There are no MZMs and EPs do not appear at zero
energy, different from the $p$-wave NHJJ discussed above. To classify
this junction properly, one EP needs to be shifted to zero energy,
which breaks particle-hole symmetry\ \citep{Ohnmacht25prl,Kawabata19prx}.
It thus leads the system falling into class A. In this case, there
is no topological invariant and the EPs are fragile. Note that the
EPs do not evolve from MZMs.

The scattering amplitudes can also be obtained accordingly. For instance,
the inelastic Andreev reflection amplitude is given by

\begin{equation}
A_{h-}=\frac{-\Delta[ZE+\Omega\sin^{2}\phi/2+iE\sin(\phi)/2]}{(Z^{2}+1)\Omega E+2ZE^{2}-\Delta^{2}(2Z+i\sin\phi)/2}.
\end{equation}
The poles of this scattering amplitude corresponds to the Andreev
quasi-bound states given in Eq.~\eqref{eq:Andreev spectrum_sp}.
The Andreev reflection probability changes smoothly as a function
of Josephson phase $\phi$. As a result, the consequent supercurrent
also varies smoothly across the EPs.

\section{Discussion and conclusion}

Our proposal of $p$-wave NHJJ is experimentally relevant. The $p$-wave
superconductor can be fabricated by 1D semiconductors such as InAs
and InSb, proximitized to $s$-wave superconductors\ \citep{Lutchyn2010,Oreg10prl,Mourik12science,Frolov20NP}.
The local non-Hermitian barrier potential at the junction could be
realized by coupling of the Josephson junction with a dissipative
lead\ \citep{ZhangS22prl,LiuD22prl}. Notably, the evolution of MZMs
to a pair of EPs should be a distinct observable signature in the
spectroscopy of a $p$-wave NHJJ.

In summary, we have investigated the exceptional Andreev spectrum
and transport properties of $p$-wave NHJJs. By solving the bulk BdG
equation, we have revealed the appearance of EPs in the spectrum of
Andreev quasi-bound states, with no counterparts in Hermitian Josephson
junctions. These EPs origin from MZMs as varying the non-Hermiticity
and are topologically protected. The supercurrent across the $p$-wave
NHJJ is obtained from the Andreev reflection amplitudes. It turns
out that the supercurrent profile shows no divergence across the EPs.
The critical supercurrent decays linearly with increasing the non-Hermiticity
strength.

\section{Acknowledgments}

We thank Fernando Dominguez for helpful discussions. This work was
supported by the Würzburg-Dresden Cluster of Excellence ct.qmat, EXC2147,
project-id 390858490, the DFG (SFB 1170).

\bibliographystyle{apsrev4-1-etal-title}

\end{document}